\documentclass[preprint,12pt£¬twocolumn]{elsarticle}
\setcitestyle{square,aysep={},yysep={;}}
\newcommand{\RNum}[1]{\uppercase\expandafter{\romannumeral #1\relax}}
\usepackage{epstopdf}
\usepackage{lineno,hyperref}
\usepackage{tabu}
\usepackage{amssymb}
\modulolinenumbers[200]
\biboptions{numbers,sort&compress}

\begin{document}

\begin{frontmatter}

\title{Robust $p$-orbital half-metallicity and high Curie-temperature in the hole-doped anisotropic $\rm{TcX_2}$ ($X=S, Se$) nanosheets}
\author[mymainaddress]{Chang-Wei Wu}
\author[mymainaddress]{Dao-Xin Yao\corref{mycorrespondingauthor}}
\cortext[mycorrespondingauthor]{Corresponding author}
\ead{yaodaox@mail.sysu.edu.cn}

\address[mymainaddress]{State Key Laboratory of Optoelectronic Materials and Technologies, School of Physics, Sun Yat-Sen University, Guangzhou, China.}

\begin{abstract}
{Here, we study the magnetism of the distorted 1$T$-\rm{$TcX_2$} ($X$=S,Se) based on first-principles calculation. The magnetism originates from the hole doping due to the density of states near the valence band edge having van Hove singularity feature. The calculated results show that the \rm{$TcS_2$} monolayer can develop an interesting ferromagnetic (FM) half-metallic phase with tunable spin-polarization orientation. The FM half-metallicity and magnetic moments of the hole-doped \rm{$TcS_2$} monolayer are primarily derived from the $p$ orbital of S atoms, then, a FM ground phase with a high Curie temperature ($T_C$) (larger than 800 K) is obtained due to the strong $S_p$-$S_p$ direct exchange interaction. The magnetic order is robust against thermal excitations at finite temperatures because of magnetic anisotropic energy. In the \rm{$TcS_2$} bilayer, the electrons near Fermi level are redistributed when introducing the interlayer interaction, which suppresses the ferromagnetism induced by hole doping. The ferromagnetism can be recovered when the interlayer interaction is weakened.}
\end{abstract}

\begin{keyword}
van Hove singularity\sep hole doping \sep half-metallicity \sep $p$ orbital\sep magnetic anisotropic energy \sep interlayer interaction
%\MSC[2010] 00-01\sep  99-00
\end{keyword}

\end{frontmatter}

\linenumbers

\section{INTRODUCTION}
Spintronics, which takes advantage of the spin degree of freedom of electrons for delivering and storing information, have attracted great interest in science and technology. It requires the precise control of magnetic properties in low dimensions with high Curie temperature ($T_C$) for fabricating spintronic devices at the nanoscale. Half metals (HMs) are a kind of important spintronic materials with one spin channel being metallic and opposite spin channel being semiconducting. They can provide the single-spin transport that is demanded in future nanoelectric devices, such as spin-polarized tips in scanning tunneling microscopy. For the practical spintronics application at the nanoscale, the two dimensional (2D) materials should possess the following important feature:(\romannumeral1) a stable ferromagnetic (FM) half-metallic ground phase; (\romannumeral2) a $T_C$ beyond room temperature. In decades, tremendous new 2D materials have been predicated theoretically and realized experimentally\cite{PhysRevLett.108.196802,1367-2630-14-3-033003,PhysRevB.88.045416,PhysRevX.3.031002,Coleman568,PhysRevLett.108.155501,10.1038/srep11512,PhysRevB.87.100401,C6NR01333C}. These materials hold tremendous potential applications in the next-generation high-performance electronics and optoelectronics\cite{doi:10.1002/aelm.201500453,Wang,Novoselov666,Lee76}. However, most of them are limited for practical applications in spintronics due to the absence of ferromagnetism. The experimental discovery of exfoliated 2D intrinsic FM materials $\rm{CrI_3}$ \cite{Huang} and $\rm{Cr_{2}Ge_{2}Te_6}$\cite{Gong} has inspired researchers to hunt for more realizable 2D magnetic materials. Although the high $T_C$ intrinsic ferromagnetism in 2D ScCl monolayer is predicated theoretically\cite{C8NH00101D} and the ferromagnetic ordering of $\rm{VSe_2}$ monolayer is maintained above room temperature\cite{Bonilla}, the magnetism originates from $d$ orbital which is not advantageous to low energy cost. Compared to the $d$-orbital counterpart, the $p$ orbital magnetic materials can have a higher Fermi velocity and longer spin coherence length due to greater delocalization of $p$ orbital, which are advantageous for high-speed and long distance spin-polarized transport. In fact, 2D $p$-orbital magnetism is available in graphene, silicene, phosphorene, etc. by many methods, such as introducing defects\cite{PhysRevB.87.195201,PhysRevB.80.075406,PhysRevLett.108.206802}, adatoms\cite{PhysRevLett.102.126807,YANG2017120,ZHANG20161373,C6CP03210A}, or edges\cite{doi:10.1021/ja710407t}. However, these modified materials are hard to synthesize experimentally due to the uncontrollability of modification. Very recently, a new class of 2D materials are found to possess novel band structure with a flat band edge. This leads to high density of states (DOS) and a sharp van Hove singularity at or near the Fermi level. The high DOS system would lead to an electronic instability, which leads to superconductivity, magnetism and other phenomena\cite{PhysRevLett.114.236602,doi:10.1021/acs.nanolett.7b00366,PhysRevLett.100.186803}. This offers a promising approach to develop controllable magnetism by carrier doping, where the carrier doping density can be controlled by gating technique. From first-principles calculations results, the GaSe\cite{PhysRevLett.114.236602}, $\rm{PtSe_2}$\cite{doi:10.1021/acs.jpcc.6b06999}, $\rm{InP_3}$\cite{doi:10.1021/jacs.7b05133}, $\rm{C_{2}N}$\cite{C7TC01399J,C6RA08254H}, the $\alpha$-SnO\cite{Seixas}, SiC\cite{WU2019306} and $\rm{PdSe_2}$ monolayer\cite{C8TC01450G} show ferromagnetism and half metallicity with carrier doping. Nevertheless, their half-metallic phases are unstable or the $T_C$ is rather low. Therefore, it remains challenging to obtain robust $p$-orbital FM half-metallicity and high $T_C$ in 2D crystals for spintronic devices.

\begin{figure}[h]
\centering
  \includegraphics[width=0.75\textwidth]{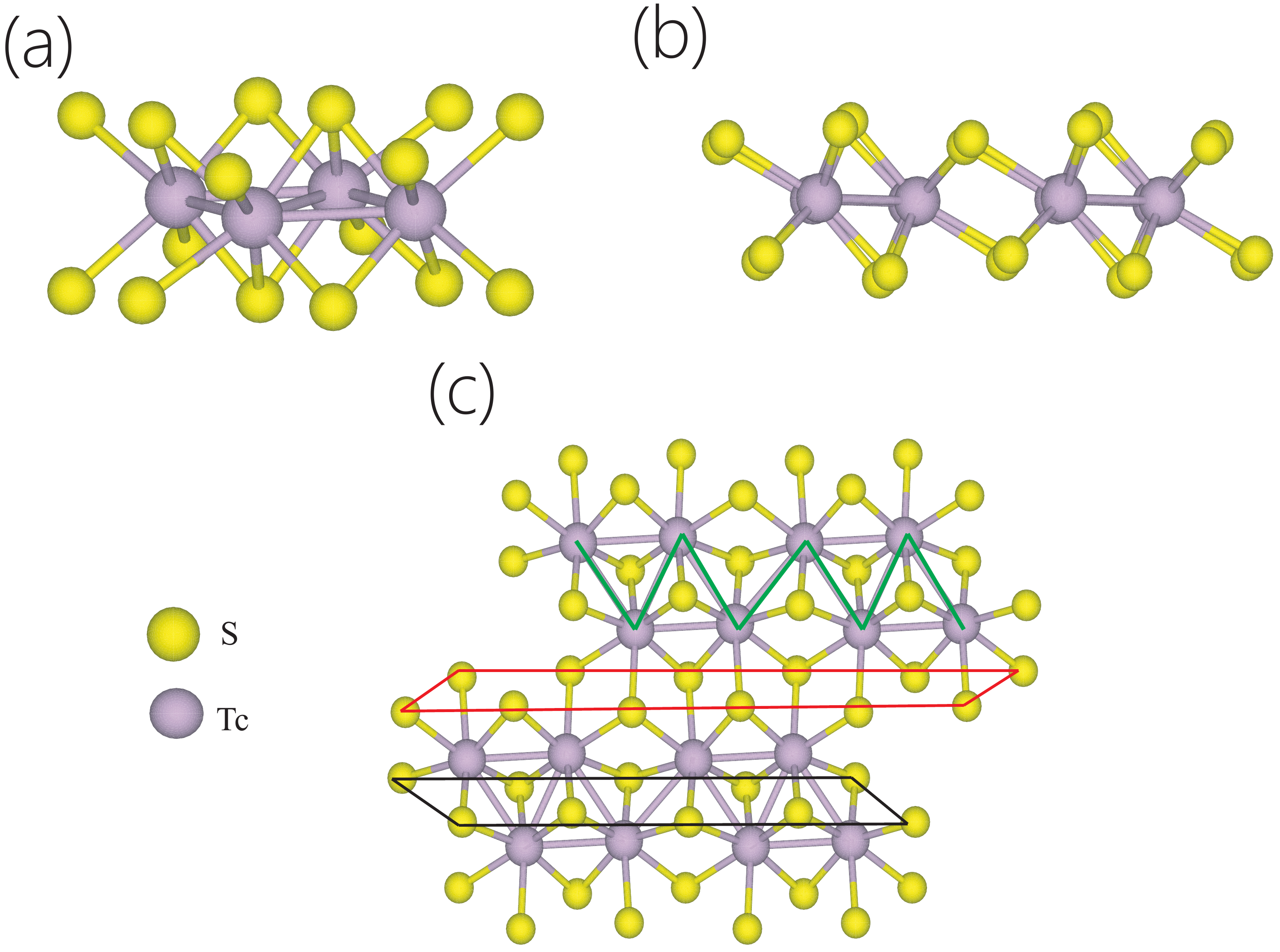}
  \caption{(a) Crystal structure of $\rm{TcS_2}$  bulk, (b) side and (c) top views of the $\rm{TcS_2}$ monolayer. The Tc diamond-shaped chains are denoted by the green zigzag line. The S atoms inter and intra the Tc chains are denoted by the red line and black line, respectively.}
  \label{fgr:Fig1}
\end{figure}

The family of transition metal dichalcogenides (TMDs) has been studied intensively because of the rich physics and tremendous potential for application, such as superconductivity\cite{Sipos} and charge density waves\cite{PhysRevLett.86.4382,doi:10.1080/00018737500101391}. Traditional TMDs adopt a sandwich-type structure where the metal atoms are located in between two layers of chalcogen atoms and exhibit isotropic behavior owing to high lattice symmetries, which determine their fundamental properties. Among the large family of TMDs, some compounds with low symmetry show more interesting anisotropic properties of both scientific and technologic importance. For example, the distorted lattice structure $\rm{WTe_2}$ displays half metallicity and large magnetoresistance effect\cite{Ali}, the single-layer $\rm{ReS_2}$ with low symmetry exhibit competitive performance with large current on/off ratios and low subthreshold swings, and a rich Raman spectrum with good signal strength\cite{Liu}. Similar to $\rm{ReS_2}$ bulk, the crystal structure of $\rm{TcS_2}$ and $\rm{TcSe_2}$ bulks are triclinic with low symmetry (space group P$\bar1$), a stable phase with distorted 1T geometry, as shown in Fig. \ref{fgr:Fig1}(a). In contrast to traditional TMDs, they possess larger and asymmetrical unit cells, corrugated surfaces, highly anisotropic structure, and metal-metal bonds. The $\rm{TcS_2}$ and $\rm{TcSe_2}$ monolayer have been predicated with extracting from bulk form using the first principle calculations, the optical properties were also reported\cite{doi:10.1021/acsami.5b12606}. However, the magnetism of $\rm{TcS_2}$ and $\rm{TcSe_2}$ monolayer have not been explored.

In this work, we systematically explore the magnetism of $\rm{TcS_2}$ system with hole doping from first-principles calculations. In the $\rm{TcS_2}$ monolayer, the system develops the FM HM over a significant range of hole doping density. More interestingly, the spin-polarization orientation of metallic band can be manipulated. Furthermore, the magnetism arises from the $p$-orbital of S atoms, which realizes strong $p$-orbital magnetism. The $T_C$ is far beyond room temperature. We also study the strain effect on magnetization, spin polarization energy, magnetic anisotropic energy (MAE), and $T_C$. On the other hand, we find that the interlayer interaction suppresses FM phase in the $\rm{TcS_2}$ bilayer. The magnetic state can be reproduced with weakening the interlayer interaction.

\section{METHODS}
Our first-principles calculations are based on the density functional theory (DFT) as implemented in the Vienna ab initio simulation package (VASP)\cite{PhysRevB.59.1758,PhysRevB.49.14251}. The ion-electron interactions is treated by projector augmented-wave (PAW) method\cite{PhysRevB.50.17953}. The electron exchange correlation potential is treated with the generalized gradient approximation (GGA) of the Perdew-Bruke-Ernzerhof (PBE) functional\cite{PhysRevLett.77.3865}. Since the band structure is central to our study, we also perform Hybrid DFT calculations based on the Heyd-Scuseria-Ernzerhof (HSE06) exchange correlation functional\cite{doi:10.1063/1.1564060}. A kinetic cutoff energy is set at 550 eV for the Kohn-Sham orbitals being expanded in the plane-wave basis. The lattice constants and the atomic positions are fully optimized with a conjugated gradient algorithm until the Hellman-Feynman forces are less than 0.01 eV/{\AA}. The interlayer interaction is described with van der Waals (VDW) forces in the DFT-D2 approximation. The convergence criterion for self-consistent field (SCF) energy is set to $10^{-6}$ eV. For Brillouin-zones integration, we employ a $\Gamma$-centered Monkhorst-Pack algorithm $16\times16\times1$ $K$-points grid and adopt the tetrahedron method with Bl\"{o}chl corrections. To calculate the MAE, we include the spin-orbit coupling (SOC) in the calculations. The energy convergence criterion is promoted to $10^{-8}$ eV, and the $K$-points grid is enlarged to $24\times24\times1$ for calculating the MAE.  In our calculation, the hole doping is simulated via changing the total number of electrons in the system, the charge neutrality is maintained by a compensating jellium background.
\section{RESULTS AND DISCUSSION}
\subsection{Electronic structure}
\begin{figure}[h]
\centering
  \includegraphics[width=0.75\textwidth]{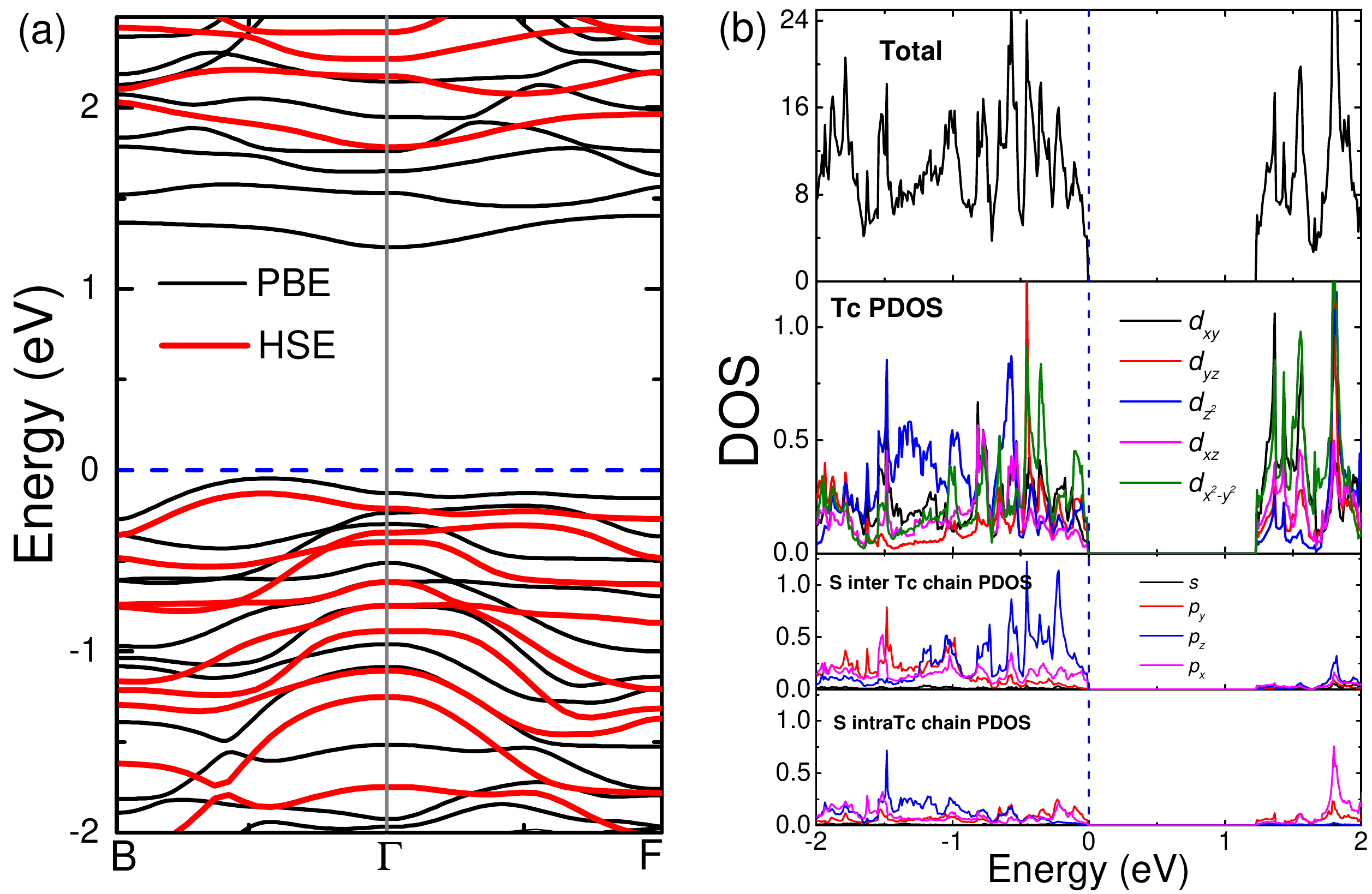}
  \caption{(a)Electronic band structure of $\rm{TcS_2}$ monolayer with PBE (black curve) and HSE (red curve). (b)Total DOS and partial DOS projected on Tc's d and C's p orbitals, where the black, red, blue, purple and green curves correspond to $d_{xy}$(s), $d_{yz}$($p_{y}$), $d_{z^2}$($p_{z}$), $d_{xz}$($p_{x}$) and $d_{x^2-y^2}$ orbitals. The Fermi level is set to zero.}
  \label{fgr:Fig2}
\end{figure}
 The optimized lattice parameters ($a$, $b$ and $c$) for $\rm{TcS_2}$ and $\rm{TcSe_2}$ bulks in our calculations are relatively smaller than those in previous DFT calculations\cite{C3DT51903A}. This is because that the van der Waals forces are considered in our calculations, which could better describe the layer-layer interaction in materials compared with previous theoretical study with the standard DFT method. The band structures of $\rm{TcS_2}$ and $\rm{TcSe_2}$ bulk obtained from PBE calculations indicate that they are both indirect semiconductors with a band gap of 1.04 and 0.86 eV, respectively. Moreover, they are slightly larger than that of previous DFT calculation (0.9 and 0.8 eV). However, the calculated structure parameters and band gap agree well with experimental values (1.0 and 0.88 eV)\cite{LAMFERS199634,PhysRevB.55.10355} (see the Table S1 and Figure S1 in the Supplemental Material). These results show that the PBE calculation is sufficient to accurately describe the structures of $\rm{TcS_2}$ and $\rm{TcSe_2}$ bulks. Given the good agreement between theory and experiment, we expect that the PBE is also suitable to investigate the electronic and magnetic property of the single-layer $\rm {TcS_2}$ and $\rm{TcSe_2}$. In the following, we take the $\rm{TcS_2}$ monolayer as a prototype to discuss its band structure and magnetic properties.

The crystal structure of $\rm{TcS_2}$ monolayer is illustrated in Figs. \ref{fgr:Fig1}(b) and \ref{fgr:Fig1}(c). The unit cell consists of two hexagonal planes of eight S atoms and an intercalated hexagonal plane of four Tc atoms bound with S atoms forming diamond-shaped chains which are derived from the metal-metal bonds. The symmetry of the monolayer crystal is reduced from the 1T structure to the distorted 1T phase. The calculated two principal $a$ and $b$ axes are $6.378$ and $6.491$ \AA, respectively. The calculated electronic band structure is displayed in Fig. \ref{fgr:Fig2}(a), which shows that the valence band maximum (VBM) is situated at midpoint along the $\Gamma$-B (0.5, 0.0, 0.0) symmetry line, while the conduction band minimum (CBM) is situated at the $\Gamma$ point. The $\rm{TcS_2}$ monolayer has an indirect band gap of 1.28eV. These values are consistent with that of previous report\citep{doi:10.1021/acsami.5b12606}. More interestingly, one band has almost flat dispersion along the $\Gamma$-F (0.0, 0.5, 0.0) symmetry line near the Fermi level. This results in a van Hove singularity in DOS near the valence band edge. We also perform HSE calculations to predict more accurate band structure (Fig. \ref{fgr:Fig2}(a)). It is observed that the band gap is larger than that of PBE, while the flat dispersion near VBM is maintained. As shown in the DOS spectra in Fig. \ref{fgr:Fig2}(b), a van Hove singularity is found below the Fermi level 0.22 eV. The van Hove singularity characteristic of DOS at or near the Fermi level would generally result in instabilities towards symmetry-breaking phase transitions, and gives rise to ferromagnetism by carrier doping\cite {PhysRevB.96.075401}. From analysis of projected density of states (PDOS), we find that the $p_z$ orbital of S atoms and the $d_{x^2-y^2}$, $d_{z^2}$ orbital of Tc atoms mainly contribute to the VBM (Fig. \ref{fgr:Fig2}(b)). The contribution of Tc $d_{x^2-y^2}$ orbital to VBM is stronger than the $d_{z^2}$ orbital. Moreover, the hybridization between them results in a $d_{z^2}$ peak at about -0.56 eV. The difference between S atoms inter and intra the $\rm{Tc}$ chains is also highlighted. It suggests that the $p_z$ orbital of DOS near the VBM comes mainly from S atoms inter the Tc diamond-shaped chains.
\subsection{Hole doping and ferromagnetism}
\label{sec:monolayer}
\begin{figure*}
 \centering
 \includegraphics[width=0.75\textwidth]{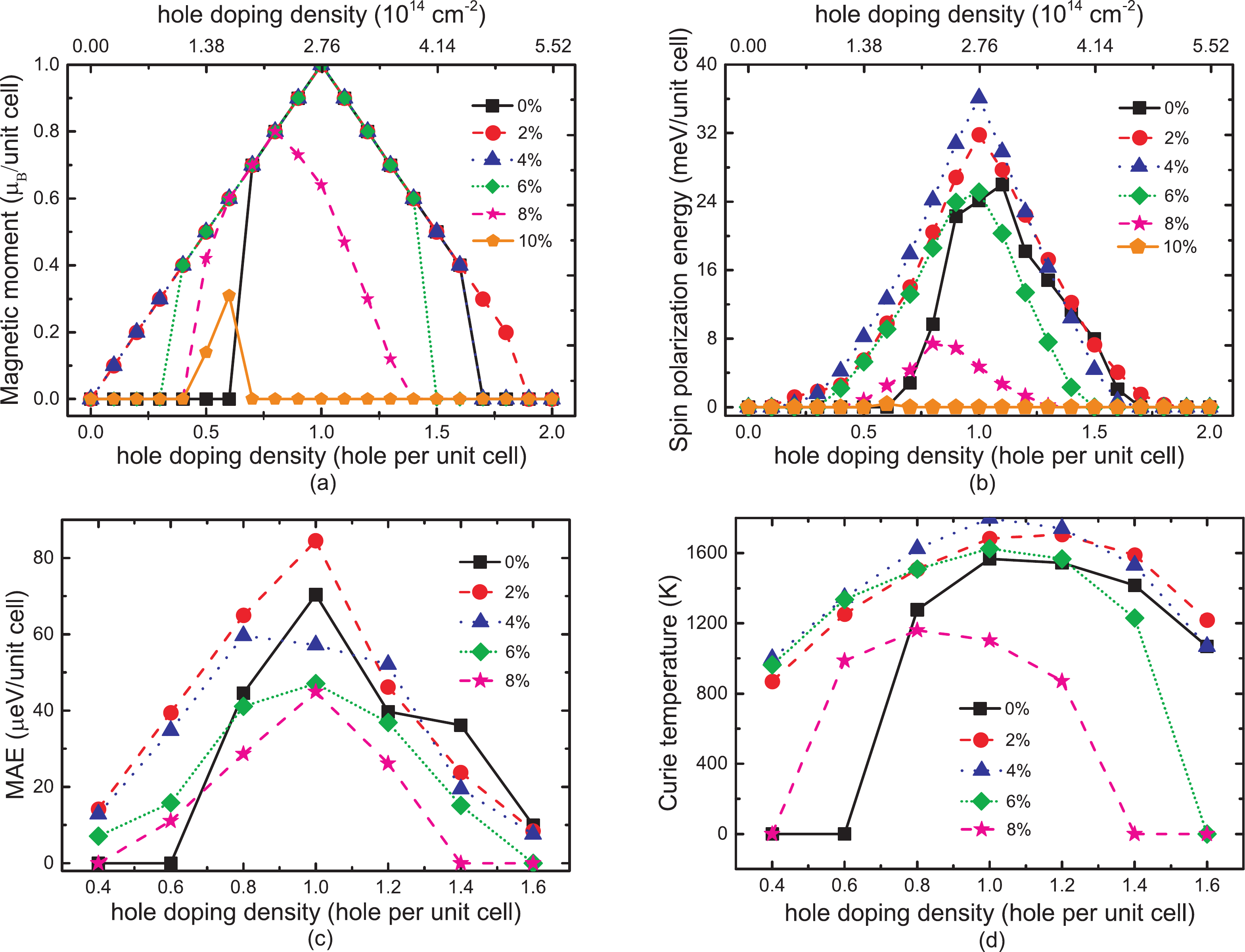}
 \caption{Magnetic moment (a), spin polarization energy (b), magnetic anisotropic energy (MAE) (c), and Curie temperature (d) with hole doping density under different biaxial strain. The black squares, red circles, blue regular triangles, green rhombus, purple stars and orange pentagon correspond to $0\%$, $2\%$, $4\%$, $6\%$, $8\%$ and $10\%$ tensile strain cases.}
 \label{fgr:Fig3}
\end{figure*}
   As mentioned above, the $\rm{TcS_2}$ monolayer is a non-magnetic semiconductor with zero doping ($\rho = 0$, where $\rho$ is the hole doping density). However, our first-principles calculations show that the system develops a spontaneous FM phase via hole doping. In Fig. \ref{fgr:Fig3}(a), we plot the magnetic moment versus the hole doping density. One clearly sees that ferromagnetism is obtained at a doping density $\rho = 1.93\times10^{14}\rm{cm^{-2}}$ (0.7 holes per unit cell). The magnetic moment of $\rm{TcS_2}$ monolayer increases linearly as a function of the hole doping density. At the doping density $\rho = 2.75\times10^{14}\rm{cm^{-2}}$ (1 hole per unit cell), the magnetic moment reaches maximum $1\rm{\mu_B}$. The magnetic moment is $1\rm{\mu_B}$/holes in a wide hole doping range from $\rho = 1.93\times10^{14}\rm{cm^{-2}}$ to $2.75\times10^{14}\rm{cm^{-2}}$, which indicates that the holes are completely spin polarized, namely the spin polarization of holes is $100\%$. However, the magnetic moment decreases linearly with continually increasing hole doping density $\rho$, and the $\rm{TcS_2}$ monolayer becomes non-magnetic phase at hole doping density $\rho$ larger than $4.416\times10^{14}\rm{cm^{-2}}$ (1.6 holes per unit cell). The large DOS not only results in magnetic instability but also gives rise to structural distortions. We relax structure by optimizing the lattice constants and the atomic positions under different hole doping density. The results (seen the Table S2 in the Supplemental Material) indicate that the relaxed structure is little different from the undoped one. We also consider the electron doping effect. The electron doping can also introduce magnetic moment, however, it fluctuates with electron doping density (see Figure S2 in the Supplemental Material). This indicates the electron doped system has no stable obvious magnetic moment compared to the hole doped system. To illustrates the stability of ferromagnetism, we calculate the spin polarization energy of the system versus the hole doping density. Here, the spin polarization energy is defined as the energy difference between the non-spin-polarized and FM phases. We also test the several antiferromagnetic phases, and we find that the antiferromagnetic phases finally converge to non-magnetic phase. Fig. \ref{fgr:Fig3}(b) shows that the polarization energy increases monotonically and it reaches the maximum (24.1 meV per hole) at the doping density $\rho =2.75\times10^{14}\rm{cm^{-2}}$, which is larger than those of $\rm{PtSe_2}$ (1.4 meV per carrier), $\rm{PdSe_2}$ (7.0 meV per carrier) and $\rm{C_2N}$ monolayer (8.5 meV per carrier), indicating its superiority for spintronics. After the doping density $\rho =2.75\times10^{14}\rm{cm^{-2}}$, the spin polarization energy decreases versus the doping density and drops to zero above $\rho =4.416\times10^{14}\rm{cm^{-2}}$.

  The physical origin of ferromagnetism can be explained by the Stoner picture\cite{Stoner1939Collective}. The system would develop ferromagnetism spontaneously when the Stoner criterion $I\times D(E_F)>1$ is satisfied, where the $I$ is Stoner interaction parameter defined as $I = {\Delta}/M$ ($\Delta$ is the spin splitting energy and $M$ is the magnetic moment) and $D(E_F)$ is the DOS at the Fermi level of the non-spin-polarized phase band structure. Hole doping increases $D(E_F)$ and may introduce magnetism due to the large DOS near the valence band edges. In our calculations, we find that the Stoner criterion is larger than 1 in a wide hole doping range from $\rho = 1.93\times10^{14}\rm{cm^{-2}}$ to $4.416\times10^{14}\rm{cm^{-2}}$. When the hole concentration is not in this range, $D(E_F)$ is not large enough and the Stoner criterion is not satisfied, thus the system is nonmagnetic. For instance, in the case of the hole doping density $\rho =2.75\times10^{14}\rm{cm^{-2}}$, the spin splitting $\Delta = 0.37$eV and the $M = 1 \rm{\mu_B}$ per unit cell, so the interaction parameter $I$ is about 0.37 eV/$\mu_B$. The required DOS at the Fermi level to satisfy the Stone criterion is $D(E_F )\geq 1/0.37\rm{eV^{-1}}$. The value of the DOS can reach $17.6\rm{eV^{-1}}$ (Figure S3 in the Supplemental Material), which is sufficiently large to develop ferromagnetism.

\begin{figure*}
 \centering
 \includegraphics[width=0.75\textwidth]{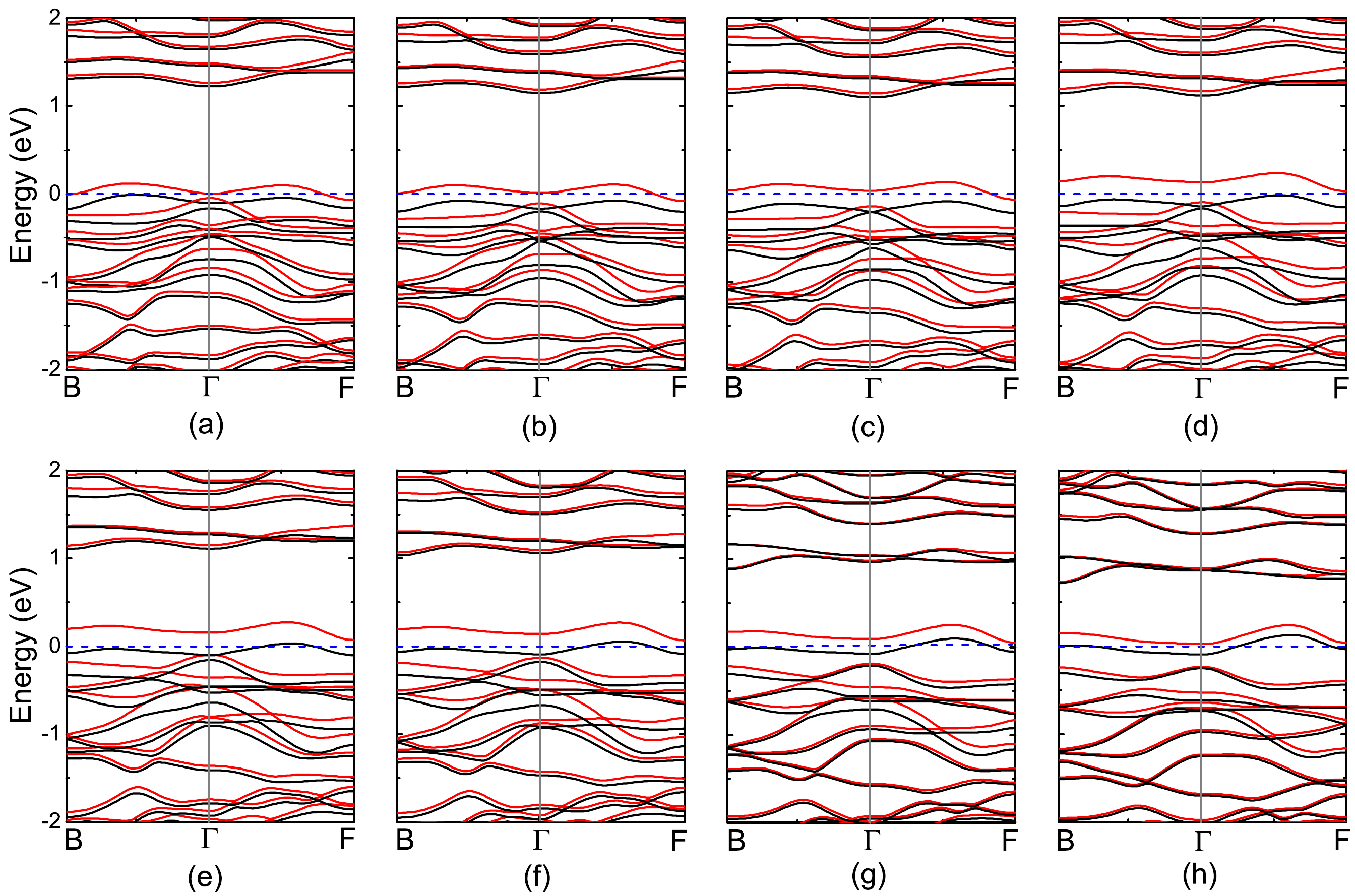}
 \caption{The band structures of $\rm{TcS_2}$ monolayer with various hole doping density, black curve for majority-spin and red curve for minority-spin. The hole density are 0.7, 0.8, 0.9, 1, 1.1, 1.2, 1.4, and 1.6 from (a) to (h), respectively. The Fermi level is set to zero.}
 \label{fgr:Fig4}
\end{figure*}
  The band structures of $\rm{TcS_2}$ monolayer under different hole doping density are displayed in Fig. \ref{fgr:Fig4}.
  The calculated results suggest that the $\rm{TcS_2}$ monolayer becomes a HM with minority-spin bands being metal characteristic and majority-spin bands being semiconductor characteristic when the hole doping density $\rho$ is larger than $1.93\times10^{14}\rm{cm^{-2}}$, as shown in Fig. \ref{fgr:Fig4}(a). The spin splitting near the Fermi level increases with increasing the hole doping density and gets to its maximum at the hole doping density $\rho =2.75\times10^{14}\rm{cm^{-2}}$ (Fig.\ref{fgr:Fig4}(d)), and the $\rm{TcS_2}$ monolayer becomes a FM spin gapless semiconductor. Subsequently, the spin splitting decreases with increasing hole doping density, and half-metallicity of the system reproduces but with an inverse spin-polarization direction (minority-spin bands exhibit semiconductor characteristic and majority-spin bands exhibit metal characteristic) (Figs. \ref{fgr:Fig4}(e)-(h)). Finally, the spin splitting vanishes when the hole doping density is larger than $4.416\times10^{14}\rm{cm^{-2}}$. The reversible spin polarization\cite{C2NR31743E,Gong8511} provides a feasible method for manipulation of spin-polarized currents and hold many advantages in spintronic application. The $\rm{TcS_2}$ monolayer shows ferromagnetism over a significant wide range of hole doping density, from $1.93\times10^{14}\rm{cm^{-2}}$ to $\rho =4.416\times10^{14}\rm{cm^{-2}}$, which is larger than that of previously reported in $\rm{C_2N}$ ($1.5\times10^{13}\rm{cm^{-2}}$ to $1.3\times10^{14}\rm{cm^{-2}}$ (0.09 to 0.78 electron per unit cell)). Moreover, the $\rm{TcS_2}$ monolayer exhibits half-metallicity over the whole range of hole doping density considered. However, half-metallicity of $\rm{C_2N}$ monolayer is observed at electron doping density from $4\times10^{13}\rm{cm^{-2}}$ to $8\times10^{13}\rm{cm^{-2}}$. The excellent half-metallicity makes $\rm{TcS_2}$ monolayer a promising candidate for spintronic applications.

  To confirm the hole-doped $\rm{TcS_2}$ monolayer is indeed HM, we also calculate the electronic structure and magnetic moment under the hole doping density $\rho =2.208\times10^{14}\rm{cm^{-2}}$ (0.8 hole per unit cell) at HSE level. It is found that the magnetic moment is also $0.8\rm{\mu_B}$, namely, the holes are fully spin polarized, which is same to that of PBE level. The DOS at HSE level (Figure S4 in the Supplemental Material) also shows the system is half-metallic phase with half-metallic gap 0.48 eV, which indicates that the half-metallicity of $\rm{TcS_2}$ monolayer could be preserved from thermally agitated spin-flip transition.

  The MAE not only reveals the ground-state magnetization orientation of system, but also indicates the relative stability of the long-range ferromagnetism against thermal excitation at finite temperatures\cite{Huang}. In this work, we calculate the MAE including the spin-orbit coupling. The MAE is defined as $MAE =E_{\bot}-E_{\|}$, where the $E_{\bot}$ or $E_{\|}$ is the total energy of hole-doped $\rm{TcS_2}$ monolayer when the easy axis is along the out-of-plane ($z$) or the lowest in plane ($x$) orientation. The MAE as a function of hole doping density is shown in Fig. \ref{fgr:Fig3}(c). It is found that the MAE can reach its maximum of 70.33 $\rm{\mu}$eV per hole for the hole doping density $\rho =2.75\times10^{14}\rm{cm^{-2}}$. This is much larger than that of $\rm {PdSe_2}$ monolayer (32 $\rm{\mu}$eV per hole)\cite{C8TC01450G}. Considering the MAE per area, the hole-doped $\rm{TcS_2}$ monolayer has a value of 3.72$\rm{\mu}eV \AA^{-2}$. The value is much smaller than that of $\rm{CoBr_2}$ (219$\rm{\mu}eV \AA^{-2}$)\cite{C7CP02158E}, but comparable to $\rm{CrXTe_3}$ (3.2-21.1$\rm{\mu}eV \AA^{-2}$)\cite{PhysRevB.92.035407} and $\rm{FeCl_2}$ (12.96$\rm{\mu}eV \AA^{-2}$)\cite{PhysRevB.92.104407}. The comparison suggests that the hole-doped $\rm{TcS_2}$ monolayer could be a useful materials in long-distance spin transport.

 Here, we use the mean field method to estimate the $T_C$ of the hole-doped $\rm{TcS_2}$ monolayer. At the mean field level, the magnetic moment can be estimated by minimizing the electronic free energy of the system at finite temperatures\cite{C8TC01450G,PhysRevLett.114.236602,2053-1583-4-2-025107}. The effect of increasing temperature is simulated by changing the Gauss smearing factor ($\sigma=K_{B}T$). We take doping density $\rho=2.75\times10^{14}cm^{-2}$ as an example to calculate the temperature-dependent magnetic moment(see Figure S5 in the Supplemental Material), which yield a critical exponent of 0.5 for the magnetic moment. By fitting the calculated magnetic moment squared to a linear function of temperature near the phase transition regime, we get an estimation of $T_C$. The calculated $T_C$ as a function of hole doping density is presented in Fig.\ref{fgr:Fig3}(d). We find that the $T_C$ is higher than 800 K when the hole doing density $\rho$ is larger than $1.93\times10^{14}\rm{cm^{-2}}$, the maximum of $T_C$ can be up to 1557 K at the hole doping density $\rho =2.75\times10^{14}\rm{cm^{-2}}$. These $T_C$ are much higher than those of GaSe (about 90 K) and $\rm{C_2N}$ monolayer (about 320 K). One should be cautions that the mean field method tends to overestimate the $T_C$. The $T_C$ values from mean field method could be modified with the empirical relation\cite{Zhifeng2017YN2} (seen Figure S6 in the Supplemental Material for more details). The modified results indicate that the $T_{C}$ is still higher room temperature.
\begin{figure}[h]
\centering
  \includegraphics[width=0.75\textwidth]{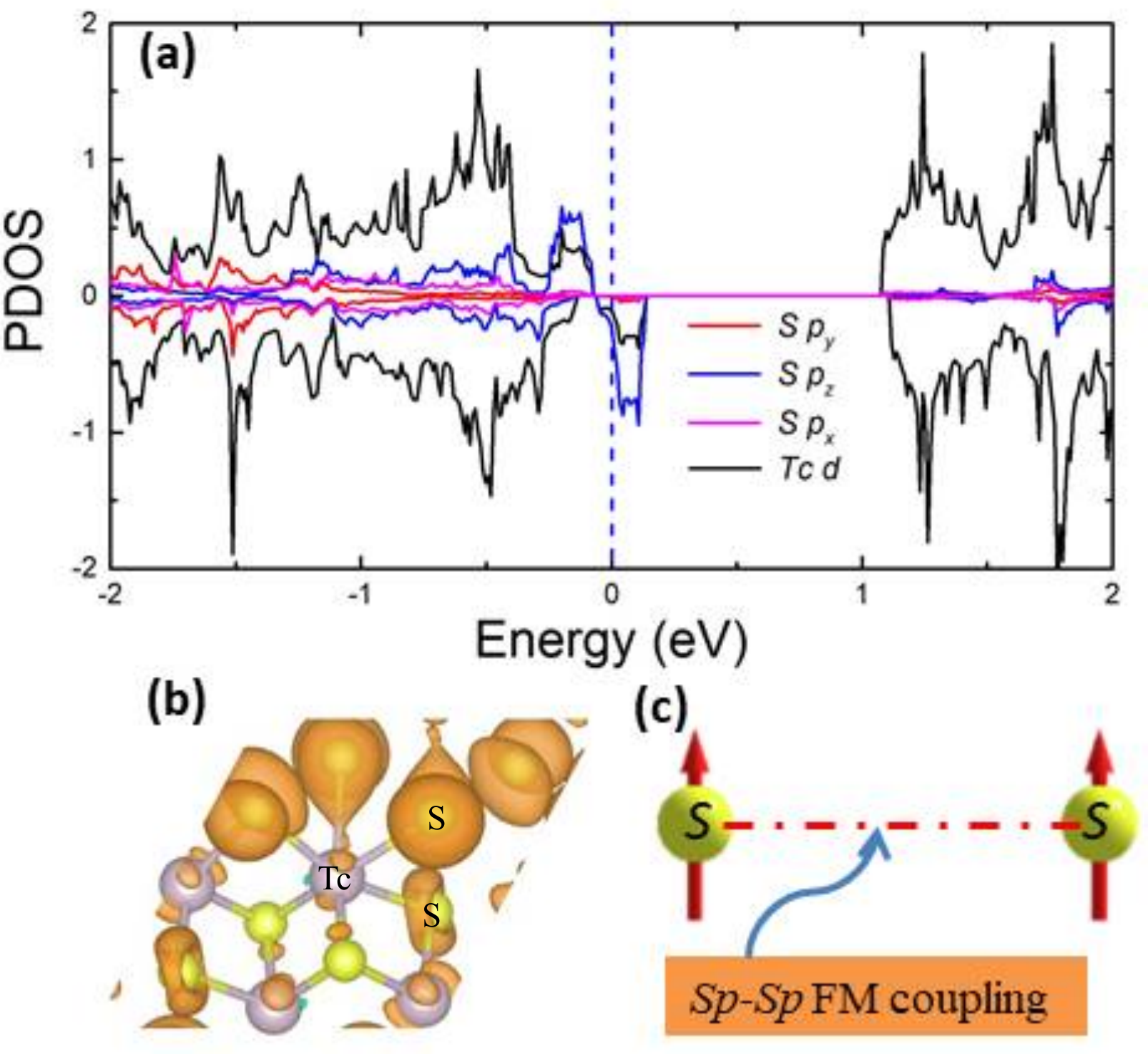}
  \caption{(a) The PDOS of the $\rm{TcS_2}$ monolayer with the hole doping $\rho=2.208\times10^{14}\rm{cm^{-2}}$, where the red, blue, purple, and black curves correspond to $p_y$, $p_z$, $p_x$ and $d$ orbitals. (b) Spin density distribution of the $\rm{TcS_2}$ monolayer with the hole doping $\rho=2.208\times10^{14}\rm{cm^{-2}}$. (c) Schematic diagram for the long-range FM coupling mechanism of $S_p$-$S_p$ direct exchange interaction.}
  \label{fgr:Fig5}
\end{figure}
 \begin{figure*}
 \centering
 \includegraphics[width=0.95\textwidth]{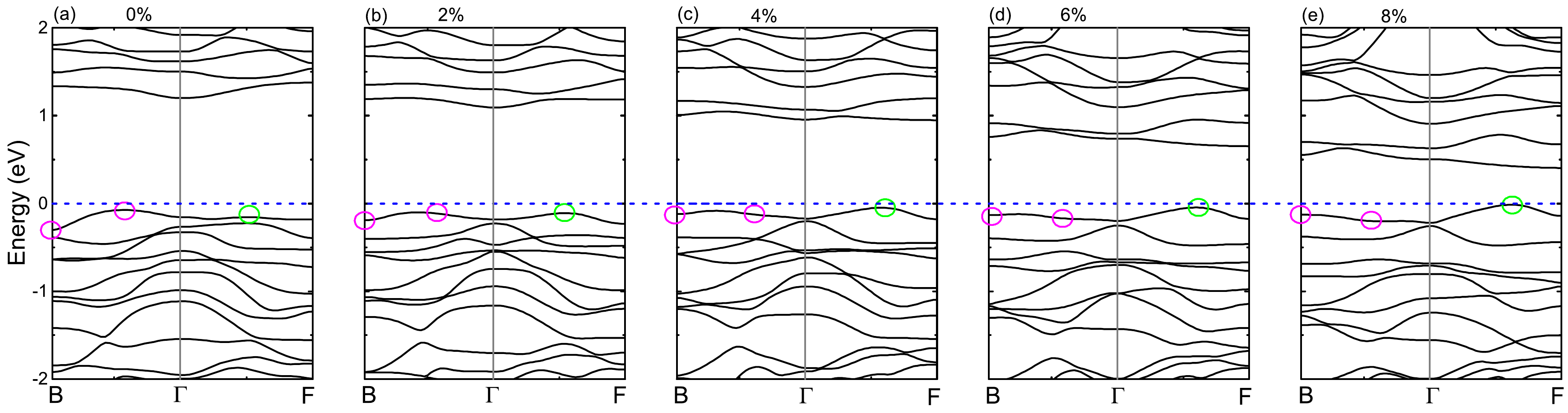}
 \caption{The band structure of the $\rm{TcS_2}$ monolayer under different biaxial strain. The Fermi level is set to zero.}
 \label{fgr:Fig6}
\end{figure*}

To clearly understand the physical origin of the high-temperature ferromagnetism and half-metallicity near the Fermi level in the hole-doped $\rm{TcS_2}$ monolayer, we take the hole doping density $\rho =2.208\times10^{14}\rm{cm^{-2}}$ as an example. The calculated PDOS (Fig. \ref{fgr:Fig5}(a)) shows that the electronic states near the Fermi level are mainly derived from the $p_z$ orbital of S atoms inter the Tc chains. Thus, it is concluded that the magnetic moment of the hole-doped $\rm{TcS_2}$ monolayer is mainly contributed by the partially occupied $p_z$ oribital of S atoms. Furthermore, the spin density distribution (Fig. \ref{fgr:Fig5}(b)) also suggests that the magnetism mainly originates from S atoms inter the Tc chains. Therefore, a strong $p$-$p$ coupling interaction is dominated around the Fermi level, which is different from the reported half-metallic one-dimensional metal-benzenetetramine ($p-d$ coupling)\cite{doi:10.1021/acs.jpcc.7b12022}. Benefiting from the spatially extended feature of S-$p$ orbitals, the $S_p$-$S_p$ direct exchange interaction (see Fig. \ref{fgr:Fig5}(c)) is extended and thus results in a strong long-range FM coupling, which can be preserved against the thermal disturbance at high temperature.

\subsection{The effect of biaxial tensile strain}
Lattice strain is also an effective method to tune the magnetic properties of 2D materials. Hence, we study the strain-dependent magnetic properties of the hole-doped $\rm{TcS_2}$ monolayer. In the present work, we only consider the biaxial tensile strain. Actually, it is reported that a critical tensile strain is larger than $12\%$ in $\rm{C_2N}$ monolayer \cite{doi:10.1063/1.4937269}and graphene, phosphorene can sustain tensile strain above $25\%$\cite{Kim,PhysRevB.90.085402}. The $\rm{TcS_2}$ monolayer can sustain a critical strain of $11\%$\cite{doi:10.1021/acsami.5b12606}. Below the point, the strained $\rm{TcS_2}$ monolayer can return to its original geometry when the tensile strain is released. In Fig. \ref{fgr:Fig3}(a), we plot the magnetic moment as a function of hole density under various biaxial tensile strain. It is observed that the hole density range for FM half-metallicity substantially increases with small tensile strain. For $2\%$ and $4\%$ tensile strain cases, the range is from 0 to $4.9\times10^{14} \rm{cm^{-2}}$ (1.8 holes per unit cell) and from 0 to $4.416\times10^{14} \rm{cm^{-2}}$, respectively, which are much larger than that of zero strain (from $1.93\times10^{14}\rm{cm^{-2}}$ to $4.416\times10^{14}\rm{cm^{-2}}$). However, the range will decrease with larger tensile strain. For the $6\%$ and $8\%$ tensile strain cases, the range is from $1.1\times10^{14} \rm{cm^{-2}}$ to $3.85\times10^{14} \rm{cm^{-2}}$ and from $1.38\times10^{14} \rm{cm^{-2}}$ to $3.58\times10^{14} \rm{cm^{-2}}$, respectively. When the tensile strain reaches $10\%$, the magnetism is almost suppressed. In Figs. \ref{fgr:Fig3}(b)-(d), we also plot the calculated spin polarization energy, MAE and $T_C$ versus hole doping density in different strain. A similar trend is observed in the maximal spin polarization energy, MAE and $T_C$ with different strain. These results imply that the strain is a powerful alternative method to tune the magnetism in the $\rm{TcS_2}$ monolayer.

In order to understand the strain-dependent magnetism, we plot the band structure of the $\rm{TcS_2}$ monolayer without hole doping under different tensile strain in Fig. \ref{fgr:Fig6}. The band extremum along the $\Gamma$-B (circled by the purple line) is gradually lowered with increasing the tensile strain, while the band at B point (circled by the purple line) is elevated gradually, and the band extremum along $\Gamma$-F (circled by the green line) is elevated gradually. This leads to the valence band along $\Gamma$-B near the Fermi level slowly becoming flat with small tensile strain, subsequently becoming dispersed with continuously increasing strain. However, the band along $\Gamma$-F always is dispersed with increasing strain. As a result, the DOS at VBM is first increased, subsequently decreased with increasing strain, then the Stoner Criterion, $I\times D(E_F)>1$, gets satisfied easily, and then satisfied hardly. According to the Stoner model, the ferromagnetism is first enhanced, subsequently reduced under hole doping, which is consistent with the discussion above.
\begin{figure}[h]
\centering
  \includegraphics[width=0.75\textwidth]{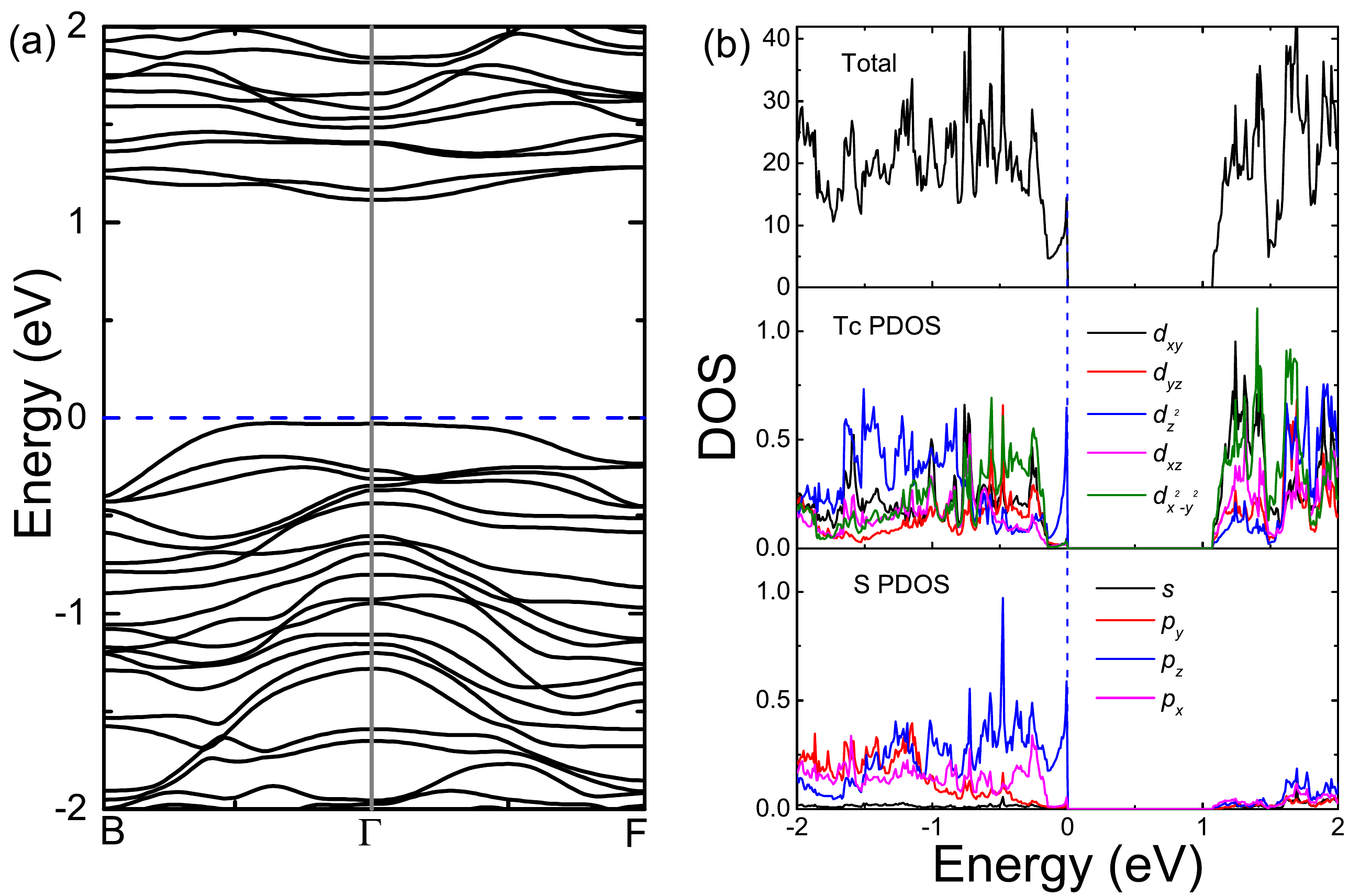}
  \caption{(a) Electronic band structure of $\rm{TcS_2}$ bilayer. (b)Total DOS and partial DOS projected on Tc's d and C's p orbitals, where the black, red, blue, purple, and green curves correspond to $d_{xy}$(s), $d_{yz}$($p_{y}$), $d_{z^2}$($p_{z}$), $d_{xz}$($p_{x}$), and $d_{x^2-y^2}$ orbitals. The Fermi level is set to zero.}
  \label{fgr:Fig7}
\end{figure}

\subsection{Magnetism of the $\rm{TcS_2}$ bilayer with hole doping}
The $\rm{TcS_2}$ bilayer is stacked along the $c$-axis with AA-stacking order. The lattice constants ($a=6.360$ and $b=6.458$ \AA) are slightly smaller than the ones calculated for the $\rm{TcS_2}$ monolayer. Specifically, the VBM of bilayer is switched to $\Gamma$ point, displaying a direct gap semiconductor, and the band gap decreases slightly to 1.13 eV, as shown in Fig. \ref{fgr:Fig7}(a), which is consistent with the previous result\cite{doi:10.1021/acsami.5b12606}. The changes indicate that the interlayer interaction plays an important role on both geometrical structure and electronic behaviors. In addition, the flat dispersion of VBM is also observed in $\rm{TcS_2}$ bilayer, and a large van Hove singularity characteristic DOS is observed at the VBM (Figs. \ref{fgr:Fig7}(a) and \ref{fgr:Fig7}(b)). Different from the $\rm{TcS_2}$ monolayer, the contribution of Tc $d_{z^2}$ orbital in the VBM dominates over Tc $d_{x^2-y^2}$ orbital in the $\rm{TcS_2}$ bilayer, and the $d_{z^2}$ peak at the energy of about -0.56 eV is vanished. Moreover, the $p_z$ peak of S atoms near the Fermi level is also shifted from -0.22 eV to -0.006 eV. These results indicate that $d_{z^2}$ of Tc atoms and $p_z$ of S atoms electrons energy is increased when the $\rm{TcS_2}$ bilayer introduces the interlayer interaction. In other words, the $d_{z^2}$ and $p_z$ electrons states become less stable with interlayer interaction.
\begin{figure}[h]
\centering
  \includegraphics[width=0.75\textwidth]{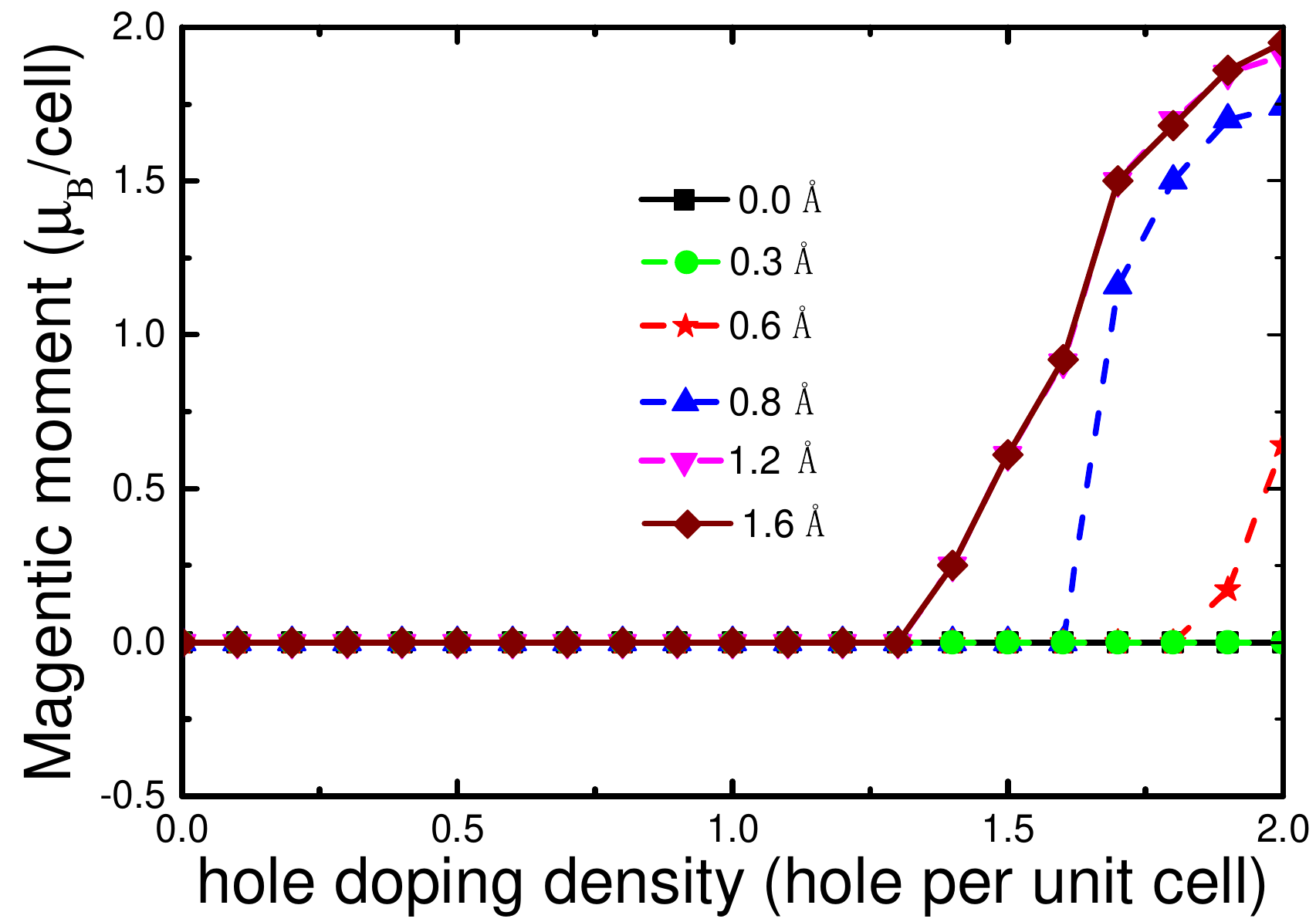}
  \caption{Magnetic moment as a function of hole doping under different expansion of layer separation. The black squares, green circles, red stars, blue regular triangles, purple inverted triangles and brown rhombus correspond to 0.0\AA, 0.3\AA, 0.6\AA, 0.8\AA, 1.2\AA and 1.6\AA layer expansion.}
  \label{fgr:Fig8}
\end{figure}
\begin{figure}
 \centering
 \includegraphics[width=0.75\textwidth]{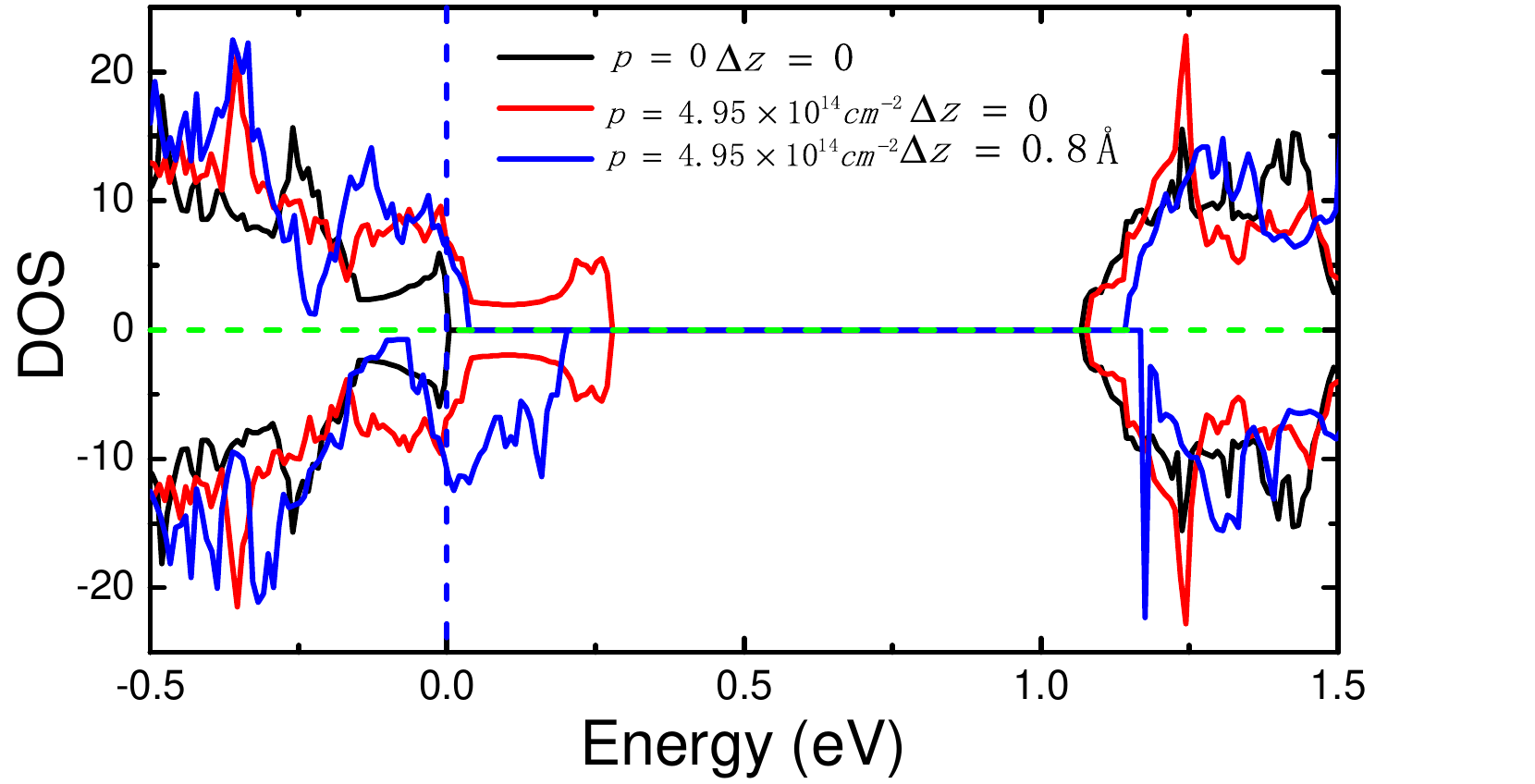}
 \caption{Spin-polarized total density of states for $\rm{TcS_2}$ bilayer. The black curve is the pristine $\rm{TcS_2}$ bilayer. The red (without layer separation expansion) and blue (with the layer separation expansion of 1\AA) curves are for the bilayer $\rm{TcS_2}$ with $\rho =4.95\times10^{14}\rm{cm^{-2}}$}
 \label{fgr:Fig9}
\end{figure}

 As suggested above, it is expected that the $\rm{TcS_2}$ bilayer would develop ferromagnetism under hole doping because of the large DOS with van Hove singularity characteristic at the energy of 0.0016 eV below the Fermi level. To examine the concept, we calculate magnetic moment as a function of hole doping density for the $\rm{TcS_2}$ bilayer and present it in Fig. \ref{fgr:Fig8}. However, the system does not exhibit magnetism under hole doping. In order to investigate whether the interlayer interaction has effect on magnetism of the $\rm{TcS_2}$ nanosheet, in Fig. \ref{fgr:Fig8}, we show the hole-dependent magnetic moment after expanding the layer separation. The magnetism appear at the layer expansion of {0.6 \AA} and the spin polarization of hole nearly reach saturation at the layer expansion of {1.6 \AA}. According to this, we conclude that the interlayer interactions have significant effect on magnetism. Since the interlayer interaction contains VDW forces and interlayer electron interaction, the two cases should be considered separately. Firstly, we test the simulation without involving VDW forces. It is found that the layer separation is shrunk by {0.8 \AA} with VDW forces in the DFT-D2 approximation. Thus, we set the layer separation equivalent to expansion of the layer separation (by {0.8 \AA}), and calculate the magnetic moment with hole doping before and after removing the DFT-D2 approximation. If VDW forces is crucial to the magnetic behavior, the magnetism should be enhanced after removing the DFT-D2 approximation. However, this do not occur in our calculation. Therefore, we conclude that the interlayer VDW forces hardly affect the formation of ferromagnetism in the $\rm{TcS_2}$ bilayer. So the interlayer electron interaction is responsible for the phenomenon.

To further study how the interlayer interaction affects the magnetic properties in the system, we plot the DOS of pristine $\rm{TcS_2}$ bilayer and that with hole doping ($\rho = 4.95\times10^{14} \rm{cm^{-2}}$, before and after layer separation expansion of 0.8\AA). As displayed in Fig. \ref{fgr:Fig9}, the hole doping (the red curves) make the VBM move towards the higher-energy level by 0.28 eV and the Fermi level crosses the valence band. On the other hand, the CBM also shifts slightly to high-energy level by 0.016 eV. However, there is no exchange splitting between the majority-spin and minority-spin channel, and the DOS curves is symmetrical. After expanding the layer separation by {0.8 \AA}, as shown in Fig. \ref{fgr:Fig9}, the hole doping (the blue curves) makes minority-spin channel shift towards high-energy level by 0.204 eV, and 0.035 eV for the majority-spin channel. The exchange splitting of the two spin channel is 0.169 eV, indicating that the breaking of time symmetry, namely resulting in reproducing ferromagnetism in the system. The reappearance of magnetism suggests that the interlayer interaction is important to develop ferromagnetism for the $\rm{TcS_2}$ nanosheets.
Based on the band-structure analysis, the flat dispersion of VBM are retained for the $\rm{TcS_2}$ bilayer. This indicates that the flat band structure (large DOS at or near the Fermi level) can not guarantee the formation of ferromagnetism for the $\rm{TcS_2}$ bilayer, namely, breaking of the time symmetry.
\begin{figure}
 \centering
 \includegraphics[width=0.75\textwidth]{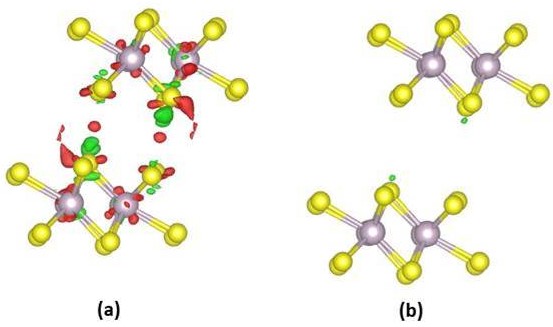}
 \caption{Charge densities difference for the $\rm{TcS_2}$ bilayer. (a) the pristine $\rm{TcS_2}$ bilayer. (b) with layer separation expansion ({0.8 \AA}). The isovalues are set at 0.002 e/{\AA}$^{3}$. The green and red color indicate deficient and accumulated electrons, respectively.}
 \label{fgr:Fig10}
\end{figure}

  We obtain the charge density difference by subtracting the single layer's charge density from that of the bilayer for the pristine bilayer and layer separation expansion ({0.8 \AA}) and display them in Fig. \ref{fgr:Fig10}. As show clearly, for the pristine bilayer, the electrons at the inner S atoms are depleted and the electrons are accumulated within the interlayer space. However, the electrons depletion and accumulation is hardly observed for the case of layer separation expansion. The phenomenon indicates that electrons transfer between layers due to the formation of multilayers. The behavior of interlayer electrons suggests that the interlayer electron interacton is closely correlated to the magnetic coupling. The hybridization of electron wave function is reduced with increasing the layer separation. Therefore, the ferromagnetism reappears upon hole doping. As discussed above, we conclude that the accumulated interlayer electrons can suppress the FM coupling in the hole-doped $\rm{TcS_2}$ bilayer. It was also reported that the ferromagnetism in the SnO bilayer depending on the balance of interlayer lone-pair electron interaction and the hole doping\cite{PhysRevApplied.8.064019}
\subsection{Discussion}
Although our work provides a general way to obtain high-temperature ferromagnetism and robust half-metallicity based on the $\rm{TcS_2}$ monolayer, it is necessary to discuss several points in the following: (\romannumeral1) The possibility to obtain $\rm{TcS_2}$ monolayer is tested by using two computational strategies\citep{doi:10.1021/acsami.5b12606}. One is to calculate the formation energy $E_f$, which determines the strength of the interlayer van der Waals interaction in the $\rm{TcS_2}$ bulk. The $E_f$ of $\rm{TcS_2}$ monolayer is estimated to 76 meV, which is smaller than that of $\rm{ReSe_2}$ monolayer (94 meV), and comparable to that of $\rm{MoS_2}$ monolayer (77 meV)\cite{doi:10.1021/jp405808a}, indicating the easy of extracting single-layer from the bulk form. The other route is to exfoliate the sigle-layer from bulk and estimate the cleavage energy $E_{cl}$. The calculated $E_{cl}$ of $\rm{TcS_2}$ monolayer is smaller that of $\rm{ReSe_2}$ monolayer and analogous to that of graphite. Given the $\rm{ReSe_2}$ monolayer and graphene have been successfully exfoliated, it is expected that the same should be successful for the $\rm{TcS_2}$ monolayer. The dynamical and thermal stabilities of the $\rm{TcS_2}$ monolayer are also confirmed by calculating phonon dispersion and ab initio molecular dynamics (AIMD) simulations with PBE functional (see Figure S7 in the Supplemental Material). (\romannumeral2) The hole doping density can be tuned by the gating technique. For instance, the carrier doping density can reach the order of $10^{14} \rm{cm^{-2}}$ in graphene by ion liquid gating\cite{PhysRevLett.105.256805}, and the order of $10^{13} \rm{cm^{-2}}$ in dichalcogenide monolayer by back-gate gating\cite{Zhang725}, even the order of $10^{15} \rm{cm^{-2}}$ by electrolyte gating\cite{doi:10.1021/ja505097m}. (\romannumeral3) In according to the Mermin-Wagner theorem, a 2D Heisenberg system does not exist long-range magnetic order at finite temperature\cite{PhysRevLett.17.1133}. However, the hypothesis of the theorem can be broken due to the magnetic anisotropy\cite{PhysRevB.38.12015} which can stabilize the magnetic order in a finite size and temperature. So it is expected that the half metallicity produced by hole doping in the $\rm{TcS_2}$ monolayer can be accessible experimentally.
\section{CONCLUSIONS}
In conclusion, we show that the $p$-orbital half-metallicity can be introduced when hole are doped into the $\rm{TcS_2}$ monolayer from first  principles calculation. The half-metallic phase with tunable spin-polarization orientation can be achieved in a wide range of hole doping density from $1.93\times10^{14} \rm{cm^{-2}}$ to $4.416\times10^{14} \rm{cm^{-2}}$. Moreover, the half-metallicity is confirmed with the HSE06 method. The strong $S_p$-$S_p$ direct exchange interaction is responsible for a FM phsae with $T_c$ above room temperature. Additionally, the lattice strain is an effective way to tune magnetism. A similar reversible modulation of magnetic moments and electronic phase transitions between half-metallic and semiconducting phases in zigzag $\rm{MoS_2}$ nanoribbons has also been revealed\cite{doi:10.1021/jz301339e}. The $\rm{TcS_2}$ monolayer is considered to be promising candidate for spintronic application due to experimental feasible synthesis, robust $p$-orbital FM half-metallicity and high $T_c$. In the $\rm{TcS_2}$ bilayer, the ferromagnetism depends on not only the hole doping density but also the interlayer interaction. In this work, we pave a way for exploring the nanoscale magnetism for spintronic applications and provide a comprehensive picture for the mechanism of the magnetic behavior in the $\rm{TcS_2}$ nanosheets.

\section*{Conflicts of interest}
There are no conflicts to declare.
\section*{Acknowledgements}
The authors thank Yijie Zeng and Wanxing Lin for helpful discussions. This project is supported by NKRDPC-2017YFA0206203, NKRDPC-2018YFA0306001, NSFC-11574404, NSFG-2015A030313176, the National Supercomputer Center in Guangzhou, Three Big Constructions-Supercomputing Application Cultivation Projects, and the Leading Talent Program of Guangdong Special Projects.
\section*{References}

\bibliography{TcS2}
\bibliographystyle{elsarticle-num}
\end{document}